\documentclass[aps,prb,twocolumn]{revtex4-1}  

\usepackage{bbm,graphicx,bm,bbm,amsmath,hyperref}   

\def\ave#1{\langle #1 \rangle}
\def\aave#1{\langle\!\langle #1 \rangle\!\rangle}
\def\cL{{\cal L}}
\def\ii{{\rm i}}
\def\sx{\sigma^{\rm x}}
\def\sy{\sigma^{\rm y}}
\def\sz{\sigma^{\rm z}}
\def\s1{{\mathbbm{1}_2}}

\def\ket#1{| {#1} \rangle}

\def\tr#1{{\rm tr}{{#1}}}

\def\etal#1{#1}

\def\tit#1{}

\begin{document}

\title{Anomalous nonequilibrium current fluctuations in the Heisenberg model}

\author{Marko \v Znidari\v c}
\affiliation{
Physics Department, Faculty of Mathematics and Physics, University of Ljubljana, 1000 Ljubljana, Slovenia}

\date{\today}

\begin{abstract}
We study fluctuation properties of a one-dimensional anisotropic Heisenberg model out of equilibrium, focusing in particular on the gapped regime. Within the open-system setting we study large-deviation properties of the spin current. Numerically evaluating the first four current cumulants in a nonequilibrium stationary state at high energies, we find that the first two cumulants scale with the system size in a diffusive way, while the 3rd and the 4th cumulants do not. This means that the model is not an ordinary diffusive spin conductor.
\end{abstract}


%05.40.-a       Fluctuation phenomena, random processes, noise, and Brownian motion
%02.50.Ga 	Markov processes 
%05.30.-d       Quantum statistical mechanics
%03.65.Yz 	Decoherence; open systems; quantum statistical methods
%05.70.Ln 	Nonequilibrium and irreversible thermodynamics
%75.10.Pq 	Spin chain models
%72.10.-d 	Theory of electronic transport; scattering mechanisms       
%05.60.Gg 	Quantum transport
%75.40.Gb 	Dynamic properties (dynamic susceptibility, spin waves, spin diffusion, dynamic scaling, etc.) 
%75.10.Pq 	Spin chain models 
%75.78.-n 	Magnetization dynamics
%72.25.-b 	Spin polarized transport (for spin polarized transport devices, see 85.75.-d)
%73.23.Ad 	Ballistic transport ; in mesoscopic

\maketitle

\section{Introduction} 
Nonequilibrium physics of quantum systems is at the forefront of today's experimental and theoretical physics because many interesting and important phenomena involve states that change in time due to the flow of currents. There are essentially two approaches to how one can theoretically study nonequilibrium systems: The first one is to start from microscopic equations of motion and try to explicitly study nonequilibrium dynamics, obtaining relations between observables of interest, while the second one is to use some macroscopic formalism that enables one to directly calculate expectation values out of equilibrium. Unfortunately, both approaches have their drawbacks. The problem with the first approach is that it is for non-trivial systems (interacting) too involved to carry it out, either because it is algebraically intractable or because it is numerically time-consuming. The problem with the second approach is that out-of-equilibrium there is no generic formalism that would be akin to equilibrium thermodynamics.

All this means that there are many open questions regarding nonequilibrium physics of interacting many-body systems. In light of that a good strategy is to first try to understand the simplest nonequilibrium phenomenon in the simplest possible system. Conceptually the simplest situation is that of a nonequilibrium steady state (NESS), in which expectation values are independent of time. A simple example is a metal rod in contact at both ends with thermal reservoirs that induce a heat flow from the hot to the cold reservoir. Such situation was in fact studied more than 200 years ago by Fourier~\cite{chaleur}, suggesting the famous Fourier's law. Fourier's law, holding for diffusive systems, was the first in a series of empirical relations between a current and the gradient of a driving potential. Understanding under which microscopic conditions are such transport laws valid is one of the unsolved problems of classical~\cite{lebowitz} as well as of quantum physics.

One of the simplest and oldest quantum models on which one could test the validity of transport laws is a one-dimensional anisotropic Heisenberg model. Suggested by Heisenberg~\cite{Heisenberg:28} (and Dirac~\cite{Dirac:29}) it describes a chain of spin-$1/2$ particles interacting via a nearest-neighbor exchange interaction. The anisotropic version, also called the XXZ model, can be written in terms of Pauli matrices $\sigma_j^{\rm x,y,z}$ as
\begin{equation}
H=\sum_{j=1}^{L-1} \sx_j \sx_{j+1}+\sy_j \sy_{j+1}+\Delta\, \sz_j \sz_{j+1}.
\label{eq:XXZ}
\end{equation}
Its usefulness goes way beyond the original proposal in the theory of magnetism: To theoreticians it serves as a paradigmatic toy model of an interacting system (via the Jordan-Wigner transformation it can be rewritten as a system of interacting spinless fermions with the interaction being proportional to $\Delta$), in mathematical physics it is a premier example of an integrable system~\cite{Bethe:31}, and it is also realized in a number of materials~\cite{spin-chain}. Studies of quantum transport in the Heisenberg model have a long history. In '90 a rigorous connection has been made between integrability and ballistic transport. Namely, using Mazur's inequality~\cite{Zotos:97} one can show that, provided the current operator has a non-zero overlap with any of the local constants of motion, transport is ballistic (i.e., non-diffusive). This might lead one to think that integrability necessarily implies ballistic transport, in particular, that the Heisenberg model is a ballistic spin conductor. However, it can happen that, due to symmetry, all overlaps with the local constants are zero in which case the Mazur inequality does not give any useful information about transport. This is exactly what happens for spin transport in the XXZ model in the zero-magnetization sector~\cite{Zotos:97}. Recently, quasilocal almost-conserved quantities have been constructed~\cite{prosen} that prove ballistic transport in the Heisenberg model at high energies for anisotropies $|\Delta|< 1$, which still leaves open the question of spin transport for $\Delta \ge 1$, being the focus of this work. 

Because of the model's integrability one could speculate that diffusive transport is unlikely as diffusion tends to be associated with some source of randomness and unpredictability, while integrable models are almost by definition as orderly as they can be. Nevertheless, various studies in recent years strongly indicate that for $\Delta>1$ and zero magnetization spin transport is in fact diffusive. This, rather controversial conjecture, is based on a finite value of the diffusion constant and a zero Drude weight~\cite{Meisner:03,Prelovsek:04,Stein:09,Langer:09,JSTAT:09,Stein:10,Affleck,Herbrych:11,Znidaric:11,Huber:12,Stein:12,Karrasch:12,Arenas:13,Stein:14,Karrasch:14}, and is obtained by many different methods, ranging from semiclassical arguments~\cite{sachdev}, linear-response calculations utilizing eigenstates~\cite{Meisner:03,Prelovsek:04,Stein:09,Stein:10,Herbrych:11,Stein:12}, or time-dependent density-matrix renormalization group (tDMRG) simulations~\cite{Affleck,Karrasch:12,Karrasch:14}, spreading of inhomogeneities~\cite{Langer:09,Karrasch:14}, or stationary nonequilibrium simulations~\cite{JSTAT:09,Znidaric:11,Arenas:13}.

In the present work we go beyond the average nonequilibrium behavior of previous works and calculate higher moments of the spin current in the Heisenberg model. Rather surprisingly, the results show that higher cumulants do not scale with system size in the same way as the first two moments. This in particular means that, even if the average current behaves in a diffusive way, scaling as $\sim 1/L$ at fixed driving, higher current cumulants do not. While we mostly focus on the scaling of current cumulants with the system size, it can be noted that they are also of intrinsic interest. For instance, higher cumulants can be experimentally measured, see e.g. Refs.~\cite{Bomze:05,Gustavsson:06,Flindt:09}, they display universal features~\cite{Flindt:09}, and can be used to do process tomography~\cite{Bruderer:14}. In certain cases of weakly-coupled systems they can be obtained in a closed form~\cite{Buca:13}. The large-deviation formalism~\cite{Touchette} that we use to calculate nonequilibrium probability distribution of the current has so-far been applied mostly to few particle quantum systems~\cite{fewparticle}, or to systems of non-interacting particles~\cite{Ates:12,Znidaric:14,Medvedyeva:13,Hurtado:13,deph:14}. Here we present calculation for an interacting many-particle system.

\section{Nonequilibrium setting} 

Nonequilibrium dynamics of the Heisenberg model shall be described in a phenomenological way by the Lindblad master equation~\cite{Breuer}. Driving is induced by two Lindblad operators that flip spins up or down at each chain end. Provided spin transport is diffusive details of the boundary driving should not matter for bulk physics. To be able to calculate fluctuation properties the Lindblad Liouvillian must be ``tilted'' by a parameter $s$, resulting in the evolution equation
\begin{equation}
\frac{{\rm d}\rho_s}{{\rm d}t}= \cL(s)\rho_s,\quad \cL(s)=\cL_0+{\rm e}^{s}\cL_++{\rm e}^{-s}\cL_-,
\label{eq:Lin}
\end{equation}
where $\cL_-(\rho_s)=2L_3\rho_s L_3^\dagger$, $\cL_+(\rho_s)=2L_4\rho_s L_4^\dagger $, $\cL_0(\rho_s)=\ii [\rho_s,H]+\sum_{j=1}^2{2L_j\rho_s L_j^\dagger} -\sum_{j=1}^4{L_j^\dagger L_j \rho_s + \rho_s L_j^\dagger L_j}$, with Lindblad operators $L_1=\sqrt{1+\mu}\,\sigma^+_1$, $L_2=\sqrt{1-\mu}\,\sigma^-_1$, $L_3=\sqrt{1-\mu}\,\sigma^+_L$, $L_4=\sqrt{1+\mu}\,\sigma^-_L$. The state $\rho_s(t \to \infty)$ is for $s=0$ equal to the NESS whose expectation value of magnetization linearly interpolates between $\pm \mu$ at chain ends, while the average current scales~\cite{JSTAT:09,Znidaric:11} as $\sim 1/L$. Introducing $N_t$ as a number of particles (magnetization) that is transferred into the right reservoir during time $t$, the average current $J_1$ is simply $J_1=\ave{N_t}/t$. In our work we would like to assess the whole probability distribution $P(J=N_t/t)$ of obtaining current $J$ in a measurement of duration $t$, and in particular evaluate cumulants of $P(J)$. The tilting parameter $s$ is introduced to facilitate that. To see how that comes about let us think in terms of a stochastic unraveling of the Lindblad equation. Namely, the solution $\rho(t)$ of the Lindblad equation can also be obtained as an ensemble average $\overline{|\psi(t)\rangle\langle \psi(t) |}$ (or, for stationary properties, as a long-time average) over stochastic wavefunctions $\ket{\psi(t)}$, also called stochastic trajectories. Each $\ket{\psi(t)}$ is probabilistically evolved either by a non-Hermitian effective Hamiltonian, or by two jump terms described by $\cL_-$ and $\cL_+$. For more details on the stochastic wavefunction approach, see, e.g.~\cite{Breuer}. When $\cL_-$ acts (i.e., $L_3$) it increases the number of particles in the chain by one, thereby decreasing $N_t$ by one, whereas when $\cL_+$ acts it increases $N_t$ by one. More precisely, one can show~\cite{Touchette} that the moment-generating function of $N_t$ is given by $\ave{{\rm e}^{s\, N_t}}=\tr{\rho_s(t)}$ (brackets here denote an ensemble average, $\ave{{\rm e}^{s\, N_t}}=\int {\rm e}^{s\, N_t} P(N_t) {\rm d}N_t$), and is therefore for long times determined by the eigenvalue $\Lambda(s)$ of $\cL(s)$ with the largest real part, resulting in 
\begin{equation}
\Lambda(s)=\lim_{t \to \infty}\frac{1}{t} \ln{\ave{{\rm e}^{s\, N_t}}}.
\label{eq:Lambda}
\end{equation} 
Once $\Lambda(s)$ is calculated, the inverse transformation can be used (the Legendre transform) to get the probability distribution. For large times one has 
%$\Phi(J)={\rm max}_s (Js-\Lambda(s))$ and $P(J) \sim {\rm e}^{-t\, \Phi(J)}$. 
\begin{equation}
P(J) \sim {\rm e}^{-t\, \Phi(J)}, \quad \Phi(J)={\rm max}_s [J\,s-\Lambda(s)].
\label{eq:Phi}
\end{equation}
The method just outlined, which can be used to calculate $P(J)$, is called the large-deviation (LD) formalism~\cite{Touchette} (in mesoscopics one often uses an imaginary parameter $\chi$, $s=\ii \chi$, calling the method full-counting statistics~\cite{Levitov}).

\section{Current distribution} 

\subsection{Illustration}
\begin{figure}[t!]
\centerline{\includegraphics[width=0.45\textwidth]{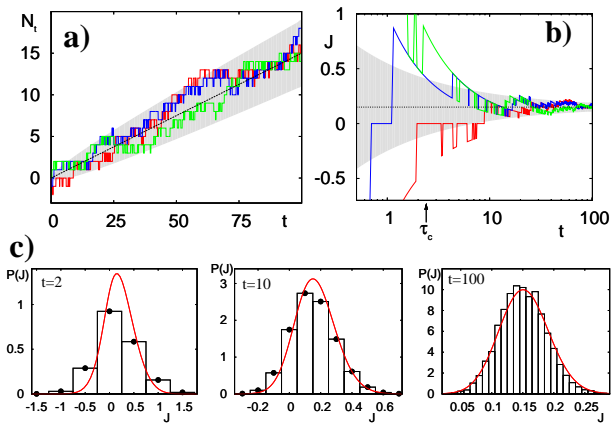}}
\caption{(Color online) Illustration of nonequilibrium current distribution: a) the number of transferred particles $N_t$ for three realizations of a stochastic process. b) The current $J=N_t/t$ for the same three realizations. Gray shading in a) and b) is theoretical long-time standard deviation obtained from the LD theory. In b) $\tau_{\rm c}\approx 2.4$ marks the convergence time towards a long-time LD behavior, the horizontal dashed line is at theoretical $J_1\approx 0.150$. c) Histograms of current distribution at three times and the theoretical prediction (full red curves). System length is $L=6$, driving $\mu=0.5$ and $\Delta=1.5$.}
\label{fig:hist}
\end{figure}
For readers not familiar with fluctuation statistics let us illustrate how $N_t$ and $P(J)$ would be measured via a stochastic wavefunction simulation, or, in an actual experiment. In Fig.~\ref{fig:hist}a is shown a number of transferred particles $N_t$ for three representative trajectories (after the stationarity is reached). One sees that $N_t$ changes by $\pm 1$ at the moments when $L_{3,4}$ is applied~\cite{foot1}. Obtaining the measured current for a particular realization via $J=N_t/t$, the plots in Fig.~\ref{fig:hist}b are obtained. Convergence to the asymptotic long-time behavior that is governed by $\Lambda(s)$ is reached after the convergence time $\tau_{\rm c}$ that is inversely proportional to the Liouville gap (for large $L$ the gap scales as $\sim 110/L^3$). Indeed, one can see that for $t \gg \tau_{\rm c}$ the average current and the standard deviation of the three samples converge to predictions of the LD theory. Making measurements of current for many trajectories, or, taking independent $N_t$ from a single long trajectory, we obtain the current distribution shown in Fig.~\ref{fig:hist}c. The distribution again converges to the theoretical one (\ref{eq:Phi}) for times larger than $\tau_{\rm c}$. Note that, due to a discreteness of $N_t$, the distribution $P(J)$ is actually a sum of delta-peaks (nicely visible for $t=2$ when $J$ can take only half-integer values), which though becomes irrelevant for large $L$ when $\tau_{\rm c}$ is large. Also, for short times there is a non-negligible probability that the measured current is actually negative (flow opposite to driving). For instance, at $t=2$ (Fig.~\ref{fig:hist}c) this probability is $\approx 0.16$, and at $t=10$ it is $\approx 0.07$.

Note that in the rest of the paper we do not calculate current cumulants via a simulated distribution function (e.g., the one in Fig.~\ref{fig:hist}c) as this would be very inefficient. As we explain in the following subsection, we calculate current cumulants directly by numerically evaluating derivatives of $\Lambda(s)$ (\ref{eq:Lambda}), i.e., using the large deviation formalism.

\subsection{Cumulants}

Our main quantity of observation is low-order current cumulants. Because we use factors ${\rm e}^{\pm s}$ in $\cL(s)$ (\ref{eq:Lin}) the current we are considering is actually a particle current with the operator at bond $k$ being $J=\sx_k \sy_{k+1}-\sy_k \sx_{k+1}$ (which differs from a true magnetization current by a trivial prefactor). From Eq.(\ref{eq:Lambda}) we see that $\Lambda(s)$ is equal to the asymptotic cumulant generating function. Therefore, current cumulants are simply obtained by taking derivatives,
\begin{equation}
\left. J_r \equiv \frac{{\rm d}^r \Lambda(s)}{{\rm d}s^r}\right|_{s=0} =\lim_{t \to \infty} \frac{1}{t}\aave{N_t^r}=\lim_{t \to \infty} t^{r-1}\aave{J^r}.
\label{eq:Jr}
\end{equation}
Double-brackets denote cumulants and we shall call $J_r$ simply a current cumulant~\cite{foot2}. Beware that $J_r$ (for $r>1$) are not equal (or related) to simpler NESS expectations of current powers, $\tr{(J^r\rho_{\rm NESS})}$, as $\aave{N_t^r}$ involves correlations of current at different times. Low cumulants can be obtained using perturbation theory in $s$, provided the zero-order operator $\cL(0)$ is solvable, which though is typically the case only in noninteracting models~\cite{Znidaric:14}, or, in small systems, e.g.~\cite{Flindt:10,Bruderer:14}. In our case, unfortunately, the only option is to numerically calculate $\Lambda(s)$. Because the size of the non-Hermitian $\cL(s)$ grows very fast as $4^L$, we used a number of different methods. For small sizes we used stochastic wavefunction simulations~\cite{Breuer} and exact diagonalization, while for larger $L$ we resorted to tDMRG~\cite{tdmrg} using a matrix-product ansatz to compactly represent $\rho_s$. We exploited symmetries by rewriting $\cL(s)$ as a non-Hermitian ladder system that conserves total ladder magnetization~\cite{deph:14} (in tDMRG adaptation for open-systems of Ref.~\cite{JSTAT:09} this also improves stability), looking for the ground state in the sector with zero magnetization. Derivatives (\ref{eq:Jr}) have been approximated by finite differences with the discretization step of $ds=0.01$ for exact diagonalization, and $ds=0.5$ for tDMRG. The latter value is a good compromise between the error due to a finite $ds$ and the truncation error of the tDMRG itself. Namely, because $\Lambda(s=0)=0$, the values of $\Lambda(s)$ are small for small $s$ and using too small $ds$ would require too large precision in the calculated $\Lambda(ds)$. For the used $ds=0.5$ we could achieve a few percent accuracy~\cite{footerr} in the calculated $J_4$, which required evaluating $\Lambda(ds)$ to five digits of precision, demanding in turn the size of $4^2 L$ matrices in the matrix-product ansatz for $\rho_s$ to be $\sim 500$. 
\begin{figure}[t!]
\centerline{\includegraphics[width=0.45\textwidth]{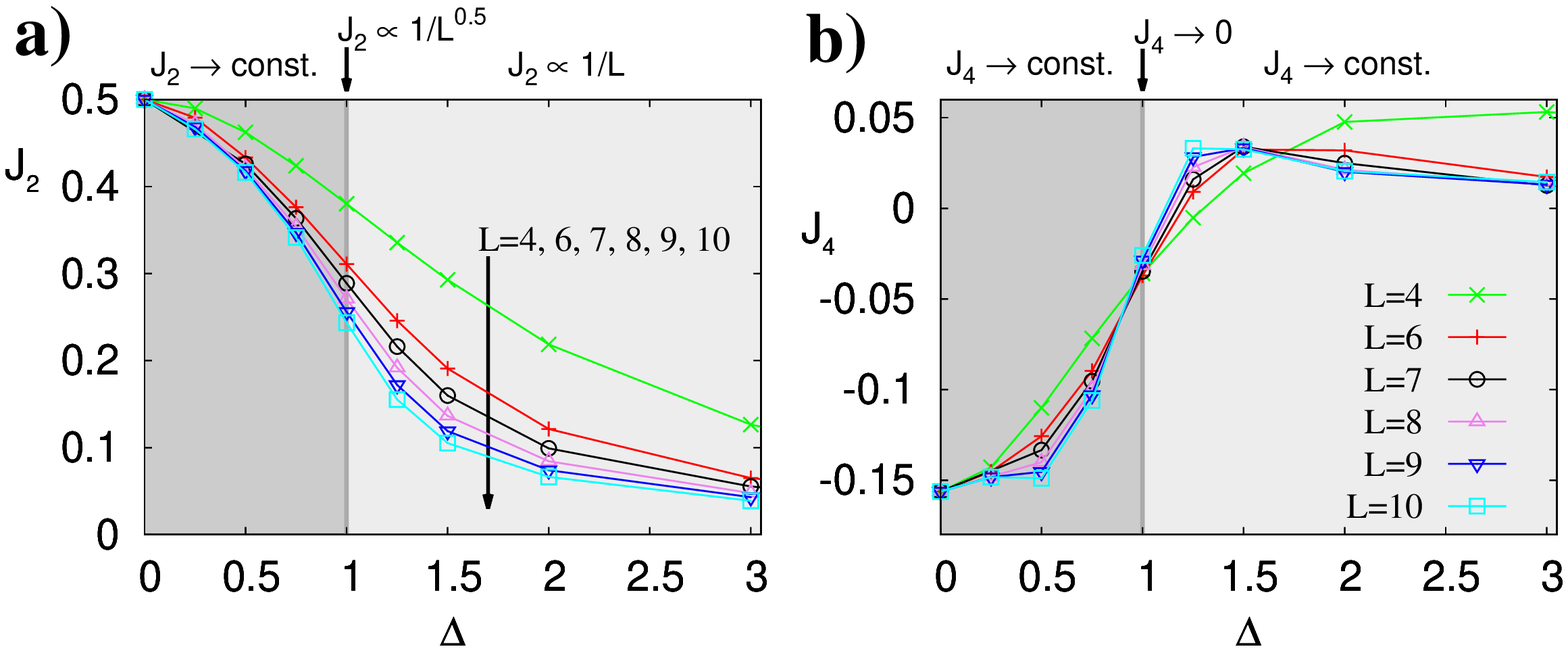}}
\centerline{\hskip2pt\includegraphics[width=0.24\textwidth]{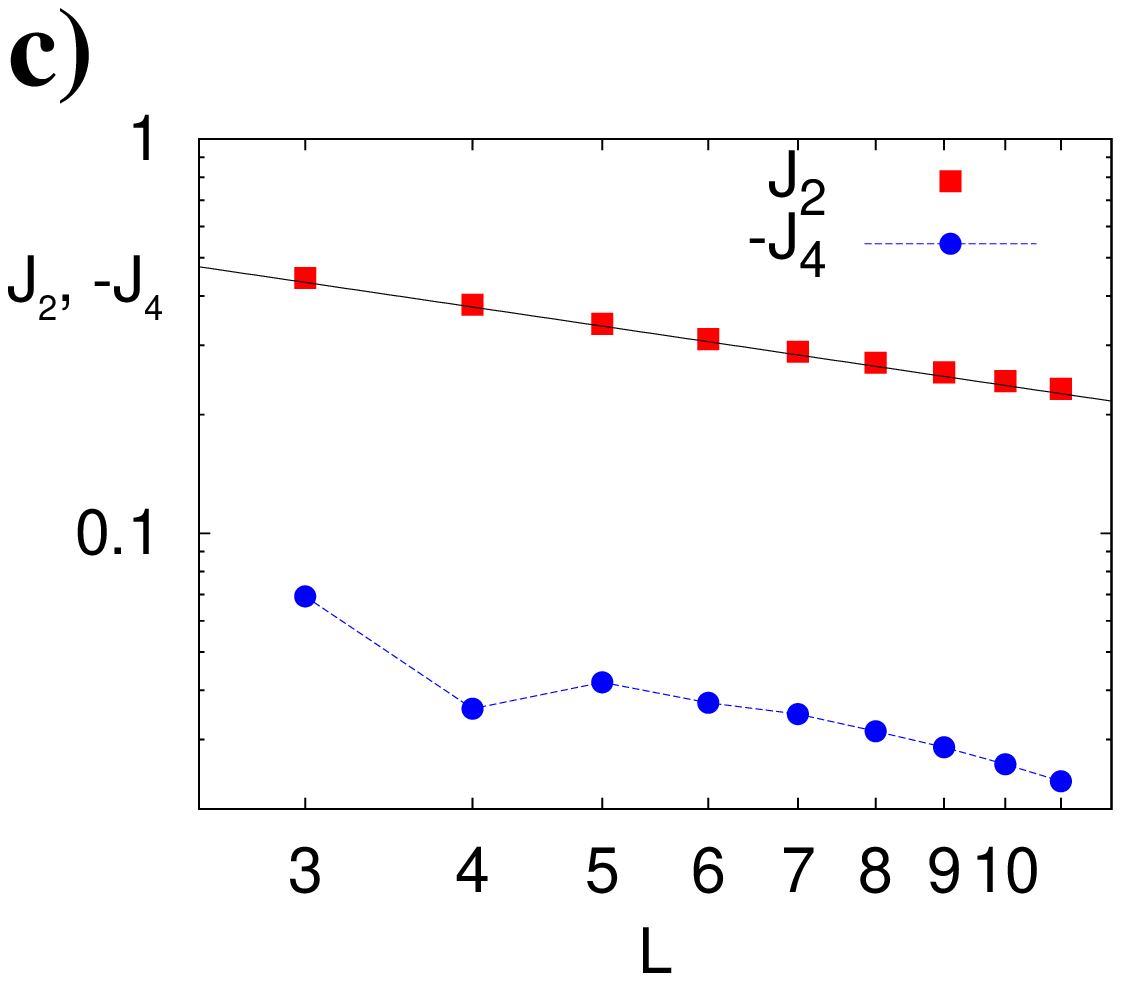}\hfill}
\caption{(Color online) Second (a) and fourth (b) current cumulants in equilibrium for different anisotropies $\Delta$. Different points are different system sizes $L$. With $\Delta$ three different phases occur, differing in the asymptotic dependence of cumulants on system size $L$. (c) Convergence of $J_{2,4}$ to zero for $\Delta=1$, the full black line indicates $J_2 \sim 0.75/\sqrt{L}$.}
\label{fig:equil}
\end{figure}

\subsection{Equilibrium}

First, we were interested in equilibrium fluctuations. For $\mu=0$ the NESS is simply $\propto \mathbbm{1}$ and is therefore an equilibrium state at infinite temperature. Due to symmetry all odd current cumulants are zero. The first two even cumulants are shown in Fig.~\ref{fig:equil}. Looking at the dependence of $J_2$ and $J_4$ on systems size $L$, keeping $\Delta$ fixed, three different behaviors are observed. For $\Delta<1$, where the model is ballistic~\cite{prosen}, $J_{2,4}$ expectedly converge to an $L$-independent value. Fluctuation properties are essentially the same as that of the XX chain~\cite{Znidaric:14}. At isotropic point $\Delta=1$ the 2nd moment scales as $J_2 \sim 1/\sqrt{L}$, while for $J_4$ we could not reliably infer the asymptotic scaling, all we can say is that $\lim_{L \to \infty}J_4=0$. In the gapped phase of $\Delta>1$ one has an interesting behavior with $J_2 \sim 1/L$, while on the other hand $J_4$ does not decay with $L$. We shall study the regime of $\Delta>1$, and in particular the scaling of cumulants with $L$, more precisely for nonequilibrium driving in the next section.

\subsection{Nonequilibrium}

\begin{figure}[ht!]
\centerline{\includegraphics[width=0.42\textwidth]{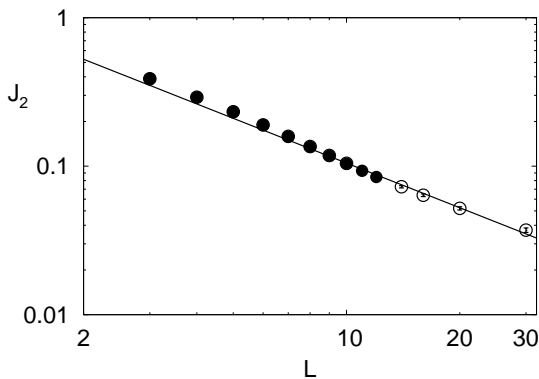}}
\caption{Scaling of a nonequilibrium current variance $J_2$ with system size $L$. Full circles are obtained by exact diagonalization, empty circles (with small error bars) are from tDMRG calculation. The full line is $\approx 1.05/L$, $\mu=0.1$, and $\Delta=1.5$.}
\label{fig:J2}
\end{figure}

\begin{figure}[ht!]
\centerline{\includegraphics[width=0.45\textwidth]{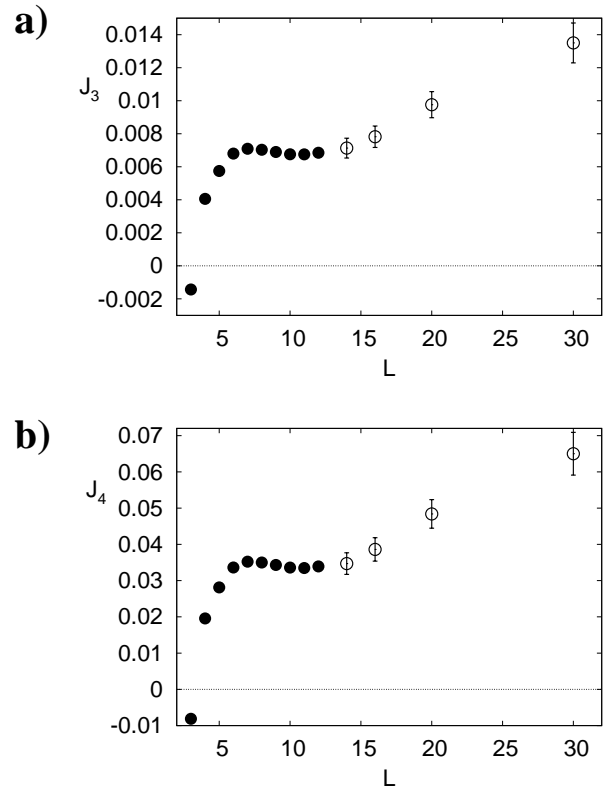}}
\caption{Scaling of a nonequilibrium current cumulant $J_3$ (a), and $J_4$ (b), with system size $L$. Full circles are obtained by exact diagonalization, empty circles (with error bars) are from tDMRG calculation. Driving is $\mu=0.1$, and $\Delta=1.5$.}
\label{fig:J34}
\end{figure}
Let us now discuss behavior out of equilibrium for $\Delta>1$, choosing a relatively small driving $\mu=0.1$. As one can see in Fig.~\ref{fig:J2} the variance scales as $J_2 \sim 1/L$. It therefore scales in the same way as the average current $J_1$ (see, e.g., data in Ref.~\cite{Znidaric:11}), with their ratio $J_1/J_2 \approx 2 \mu $ being independent of $L$. 

While it might not be a priori clear what scaling of higher cumulants one should expect in a diffusive system, an operational ``definition'' of a diffusive behavior could be that of a stochastic random-walk-like model. In a classical stochastic boundary-driven diffusive system in which particles in bulk randomly jump to the left and right with equal probability -- the so-called symmetric simple exclusion process -- all current cumulants scale as~\cite{derrida:04} $\sim 1/L$. Similar is the case for other diffusive classical~\cite{diff} and quantum~\cite{deph:14} systems, as well as for a simplistic 2-state Markov model of diffusion (see the Appendix). 

In the XXZ model the first two current cumulants therefore scale as expected for a diffusive system, namely, as $J_1 \approx 2.0\mu/L$ and $J_2 \approx 1.0/L$ (for $\Delta=1.5$). Going though to the 3rd and 4th cumulant, shown in Fig.~\ref{fig:J34}, a very different behavior is observed. While the asymptotic behavior for large $L$ is hard to predict from our finite-$L$ data, they certainly do not decay with the system size as $\sim 1/L$ and therefore do not scale in a diffusive way. While for $L \le 30$ they are in fact still slightly increasing, they can not grow with $L$ asymptotically~\cite{footL}. Considering that, and the small gradient at $L \approx 30$, (much smaller than in $J_{1,2}$) a plausible conjecture would be that $J_{3,4}$ eventually converge to an $L$-independent asymptotic values. Such behavior would be typical for a ballistic system. Worth mentioning is also that, regardless of the asymptotic behavior of $J_{3,4}$, data in Figs.~\ref{fig:J2} and~\ref{fig:J34} show that the convergence length-scale of $J_{1,2}$ and of $J_{3,4}$ is different, e.g., $J_{1,2}$ clearly converge already for $L \ll 30$. Behavior of higher current cumulants therefore reveals new physical behavior not discernible in the average current. In summary, current cumulants for the gapped XXZ model behave in an anomalous way -- the first two scale diffusively with system length while the 3rd and 4th do not. 

Note that recently some other odd behavior, not compatible with standard diffusion, has been observed. In the gapped phase there are long-range nonequilibrium correlations whose size does not decay~\cite{Prosen:10} as $\sim 1/L$ (as would be usual for a diffusive nonequilibrium system). Also, the dependence of the current on $\Delta$ seems to depend on the precise setting~\cite{Karrasch:14,Stein:10,Znidaric:11}, and there seems to be an anomalous behavior for small wavevectors~\cite{Stein:12}.

\section{Conclusion}

Using large-deviation formalism we have studied equilibrium and nonequilibrium spin current fluctuations in a one-dimensional Heisenberg model, a paradigmatic model of strongly interacting many-body systems. Using numerical methods we have calculated the first four cumulants and shown that, while the first two cumulants are inversely proportional to system length, the 3rd and 4th are not. Higher cumulants therefore do not scale as one would expect for a diffusive system. This shows that the integrable XXZ spin chain in the gapped phase at half-filling is not an ordinary diffusive conductor. 

The author acknowledges discussions with Toma\v z Prosen and support by Grant No. P1-0044 of the Slovenian Research Agency (ARRS).

%2780 besed, 560 slik, 80 enacb=3440
% hist.eps: 672x472=1.42=127 ; equil.eps: 672x287=2.34=85 ; j2.eps: 362x253=1.43=125 ; j34.eps : 358x470=0.76=220 ---> vse=560
% 5 math = 5x16=80

\section*{Appendix}

\subsection{A $2$-state Markov model}
\begin{figure}[ht!]
\centerline{\includegraphics[width=0.2\textwidth]{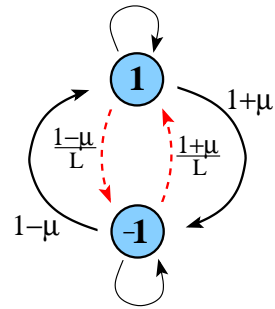}}
\caption{A minimalistic Markov model of diffusion. Full vertical arrows denote transition rates due to a coupling with a bath (Lindblad operators $L_3$ and $L_4$), while dashed arrows describe in an effective way the effect of diffusive $\sim 1/L$ current. Parameter $\mu$ is the driving strength.}
\label{fig:diagram}
\end{figure}

Let us illustrate $\sim 1/L$ scaling of all current cumulants in a simple model. In a diffusive system the restoring current scales as $\sim 1/L$ with system size. A minimalistic way to model a boundary driven spin chain would be to limit description just to the boundary spin at the right chain end, where we count the transferred particles. Introducing two states describing the boundary spin being either up or down, the two Lindblad operators $L_{3,4}$ flip the state with rates $1\pm \mu$. In addition to bath-induced flips, there is also a possibility that the spin gets flipped due to a restoring current of size $\sim 1/L$. A two-state Markovian description of such a model is depicted in Fig.~\ref{fig:diagram}.
Because $N_t$ counts the number of transferred particles into the bath (i.e., operations of $L_{3,4}$), the tilted Markov rate matrix is
\begin{equation}
\cL(s)=\begin{pmatrix}
-(1+\mu)-\frac{1-\mu}{L} & (1-\mu){\rm e}^{-s}+\frac{1+\mu}{L} \\
(1+\mu){\rm e}^{s}+\frac{1-\mu}{L}  & -(1-\mu)-\frac{1+\mu}{L}
\end{pmatrix}.
\end{equation}
The largest eigenvalue $\Lambda(s)$ can be easily calculated and the cumulants $J_r$ evaluated (\ref{eq:Jr}). For large $L$ and small driving $\mu$ the leading order expressions are $J_{2r+1}\asymp \frac{2\mu}{L}$ for odd orders, and $J_{2r}\asymp \frac{1}{L}$ for even orders. 
%\begin{equation}
%J_r=\begin{cases}
%\frac{2\mu}{L}, & r\mbox{ odd}\\
%\frac{1}{L}, & r\mbox{ even}\\
%\end{cases}.
%\end{equation}
All cumulants therefore scale as $\sim 1/L$. The ratio of the first two is $J_1/J_2=2\mu$, which is approximately the same value as obtained for the anisotropic Heisenberg model.


\begin{thebibliography}{}
\bibitem{foot1} The fact that $N_t$ can change only by $\pm 1$ is a consequence of a dynamical symmetry $U=\exp{(-\ii \alpha \sum_j \sz_j)}$. Because $U\cL(\rho)U^\dagger=\cL(U\rho U^\dagger)$, nondegenerate eigenstates of $\cL$ are block-diagonal in the eigenbasis of the total magnetization~\cite{Buca:12}. For a stochastic wavefunction this in particular means that after long time $\ket{\psi(t)}$ is an eigenstate of the total magnetization, and therefore Lindblad operators change $N_t$ by $\pm 1$.
\bibitem{foot2} Regardless of a trivial $t^{r-1}$ prefactor comming simply because we are studying $J$ and not $N_t$. Observe that this prefactor is e.g. responsible for a $\sim 1/\sqrt{t}$ decreasing standard deviation in Fig.~\ref{fig:hist}b.
\bibitem{footerr} For the largest $L=30$ the total estimated error in $J_4$ (Fig.~\ref{fig:J34}) is $\approx 9\,\%$, $6\,\%$ due to tDMRG truncation, and $3\,\%$ due to finite difference with $ds=0.5$. Finite-difference error is estimated in two ways: by directly comparing data for smaller $L$ between $ds=0.01$ and $ds=0.5$, and by using the known theoretical error of finite-difference approximation.
\bibitem{footL} Local particle current operator $J$ is a bounded operator (the largest eigenvalue is $2$) and therefore, asymptotically for large $L$, it can scale at most as $J \propto L^0$ (as is the case for ballistic transport). In particular, we can not have $J \propto L$. A similar argument holds for higher cumulants.
\end{thebibliography}
\end{document}